\documentclass[apjl]{emulateapj}

\slugcomment{}

\shorttitle{The first polarimetric signatures of infrared jets in X-ray binaries}
\shortauthors{Shahbaz et al.}

\begin{document}

\title{The first polarimetric signatures of infrared jets in X-ray binaries}

\author{T. Shahbaz\altaffilmark{1,2},
        R.P. Fender\altaffilmark{3},
        C.A. Watson\altaffilmark{4},
        K. O'Brien\altaffilmark{5}
}

\altaffiltext{1}{Instituto de Astrof\'\i{}sica de Canarias, 38200 La Laguna,
Tenerife, Spain}
\altaffiltext{2}{tsh@iac.es}
\altaffiltext{3}{School of Physics and Astronomy,
The University of Southampton, Southampton, SO17 1BJ, UK}
\altaffiltext{4}{Department of Physics and Astronomy, University of Sheffield,
Sheffield, S3 7RH, England}
\altaffiltext{5}{European Southern Observatory, Alonso de Corovo, Santiago, 
Chile}
	
\begin{abstract}
We present near-infrared linear spectropolarimetry of a sample of persistent
X-ray binaries, Sco\,X--1, Cyg\,X--2 and GRS\,1915+105. 
The slopes of the spectra are shallower than what is expected
from a standard steady-state accretion disc, and can be 
explained if the near-infrared flux contains a
contribution from an optically thin jet.
For the neutron star
systems, Sco\,X--1 and Cyg\,X--2, the polarization levels at 2.4$\mu$m are
1.3$\pm$0.10\% and 5.4$\pm0.7$\% respectively which  is greater than the
polarization level at 1.65$\mu$m. This  cannot be explained by interstellar
polarization or electron scattering in the anisotropic environment of the
accretion flow. We propose that the most likely explanation is that this is the
polarimetric signature of synchrotron emission arising from close to the base
of the jets in these systems.  In the black hole system GRS\,1915+105 the
observed polarization, although high (5.0$\pm1.2$\% at 2.4\,$\mu$m), may be
consistent with interstellar polarization.  
For Sco\,X--1 the position angle of the radio jet  on the sky is approximately 
perpendicular to the near-infrared position angle (electric vector), suggesting that 
the  magnetic field 
is aligned with the jet. These observations may be a first
step towards probing the ordering, alignment and variability of the outflow magnetic
field in a region closer to the central accreting object than is
observed in the radio band.

\end{abstract}


\section{Introduction}
\label{intro}

In the past decade or so overwhelming evidence has pointed to a clear coupling
between accretion and the formation of relativistic jets in galactic X-ray
binary systems (see \citealt{Fender06}).  Accretion states associated with hard
X-ray spectra appear to be associated with the production of a relatively
steady, continuously replenished and partially self-absorbed outflow
\citep{Fender01}, while major outbursts are associated with more discrete
ejection events which may be resolved and tracked with radio interferometers
(e.g. \citealt{Mirabel94}). 

In the radio band the steady jets observed during the hard X-ray states have a
flat ($\alpha \sim 0$, where $S_{\nu} \propto \nu^{\alpha}$) spectrum, probably
resulting from self-absorption in a self-similar outflow (e.g.
\citealt{Blandford79}). Above some frequency this flat spectral component
should break to an optically thin spectrum ($\alpha \sim -0.6$) corresponding
to the point at which the entire jet becomes transparent. There is some
evidence from some black hole X-ray binaries that this break occurs around the
near-infrared spectral region (e.g.  \citealt{Corbel02}; \citealt{Nowak05};
\citealt{Homan05}; \citealt{Russell06}),  something which can be well fit by
jet models  (\citealt{Markoff01}; \citealt{Markoff03}).  Thus the case is
strong that there is a significant contribution of synchrotron emission,
probably optically thin, in the near-infrared spectral regimes of X-ray binaries.
However, one key test which is yet to be reported is a measurement of the
linear polarization in this regime. Not only would a high level of linear
polarization confirm the synchrotron interpretation, it would offer us the
opportunity to study the degree of ordering and orientation of the magnetic
field at the base of the jet. In this paper we present near-infrared
spectropolarimetry for a sample of luminous X-ray binaries, all of which can be
confidently identify as jet sources based on radio observations.

\section{Observations and data reduction}

We obtained HK spectropolarimetry of three X-ray binaries Sco\,X--1, Cyg\,X--2
and GRS\,1915+105 with  UKIRT during the nights of 18, 19, 22 and 23 July 2004.
UIST  was used with the HK (1.4$\mu$m--2.5$\mu$m) grism and the  InSb array
with a 2 pixel wide slit, giving a spectral resolution of 680\,km\,s$^{-1}$ at
2.2$\mu$m. We used the IRPOL2 polarimetry module which comprises of a
half-wave retarder, a focal plane mask and a Wollaston prism which splits the
light from each aperture mask into orthogonally polarized ordinary (o-) and
extraordinary (e-) beams which are then projected onto the array. Rotating the
half-wave retarder to 0.0$^{\circ}$, 45$^{\circ}$, 22.5$^{\circ}$ and
67.5$^{\circ}$ allowed us to obtained spectropolarimetry. The normalized Stokes
parameters are then determined from ratios of the two intensities measured in
each frame, canceling any changes in sky transmission between images.

To determine the polarization spectrum we took short exposures of the target
and sky pair (obtained by nodding the target along the slit) at the four
half-wave retarder   positions giving rise to 8 spectra of the target and a
measurement of the polarization state of the target.  The target spectra were
proceeded by observations of an F or A-star, which were used to remove the
telluric atmospheric features. Furthermore, observations of a non-polarized
object and a polarized object were observed to determine the instrumental
polarization level and positional angle. A journal of observations is presented
in Table\,\ref{TABLE:LOG}

The ORAC-DR pipeline was used to reduce the data, extract the spectra and
determine the polarization state of the targets. First the bad pixel mask was
applied and then the images were dark-subtracted, flat-field and
sky-subtracted. The o- and e- beams were then optimally extracted 
(which also allows for any slight spectral tilt)
and wavelength calibrated using the Argon arc (rms error of 1.7$\AA$). 
\citet{Lazzati03} have noted that the values of the shape of the  polarization
spectrum  can depend on the the extraction technique.  We used different
methods (optimal, normal extraction and tracing) and find no such dependancy.
POLPACK was
then used to determine the intensity $I$ and the Stokes $Q$ and $U$ vectors.
The flux calibration was done by using the telluric star. The target spectra
were divided by the telluric star  (with the stellar features interpolated),
and then multiplied by the flux of the telluric star, determined from a
blackbody function with the same effective temperature and flux as the telluric
star.

\section{Spectra}

\begin{figure*}
\includegraphics[angle=0,scale=.80]{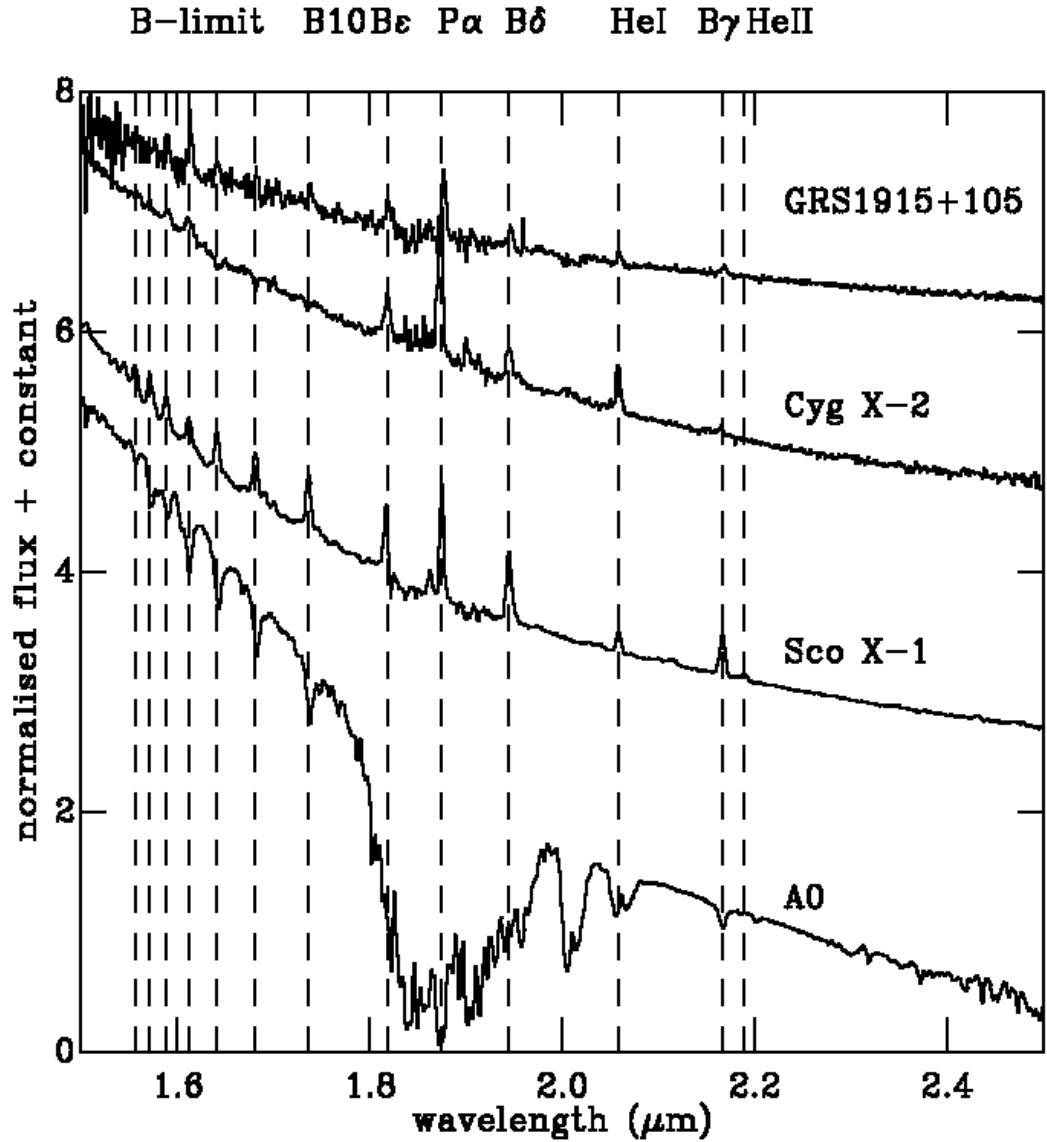}
\caption{
The dereddened HK spectrum of Sco\,X--1, Cyg\,X--2 and  GRS\,1915+105.  The 
spectrum of a telluric A0-star is also shown.  The spectra have been normalised
by dividing by the flux at 2.25\,$\mu $m and then offset on the y-axis by
adding a multiple of 1 to each spectrum.}
\label{FIG:DERED}
\end{figure*}

In Figure\,\ref{FIG:DERED} we show the dereddened HK spectra of Sco\,X--1,
Cyg\,X--2 and  GRS\,1915+105  using colour excess $E(B-V)$  values of 0.3
\citep{Vrtilek91},  0.4 \citep{Orosz99} and 6.3  \citep{Chapuis04} respectively
and the extinction law from \citet{Howarth83}.  
The $H$ and $K$ slit-magnitudes obtained after integrating the spectra with the
filter responses are as follows; $H$=11.6, $K$=11.5 for Sco\,X--1, $H$=13.2,
$K$=13.0 for Cyg\,X--2 and  $H$=11.7, $K$=11.4 for GRS\,1915+105.
The dereddened spectra can be
described as a power-law ($F_{\nu} \propto \nu ^{\alpha}$) with indicies of
1.46$\pm0.04$, 0.98$\pm0.02$ and 1.21$\pm0.05$ for Sco\,X--1, Cyg\,X--2 and
GRS\,1915+105 respectively. 
The slopes  are not consistent with what is expected
from a standard steady-state accretion disc (index of 1.5; \citealt{Bandy99}). 
Indeed the slopes
are shallower and can be explained if the near-infrared flux contains a
contribution from an optically thin jet.

\section{Polarization}
\label{sec:poln}

For each target we obtain eight spectra  for each single set of
observations of a target star. At each of the four wave-plate angles,
we measure the flux in the two orthogonal (o- and e-) polarizations. 
These measurements are combined to produce the normalized Stokes
$q$ (=$Q/I$) and $u$ (=$U/I$) spectra.   The polarization  spectrum $p$
and position angle $\theta$ is then be calculated using $p= \sqrt{q^2
+ u^2}$ and  $\theta=0.5\tan^{-1}(u/q)$ respectively. The Stokes
parameters may contain some component of instrumental polarization
introduced by reflections within the telescope and detector optics. 
The instrumental polarization was determined from  the
non-polarized standard star observed each time the science target was 
observed, and was typically $<0.1$\%. 

\begin{figure*}
\includegraphics[angle=-90,scale=.80]{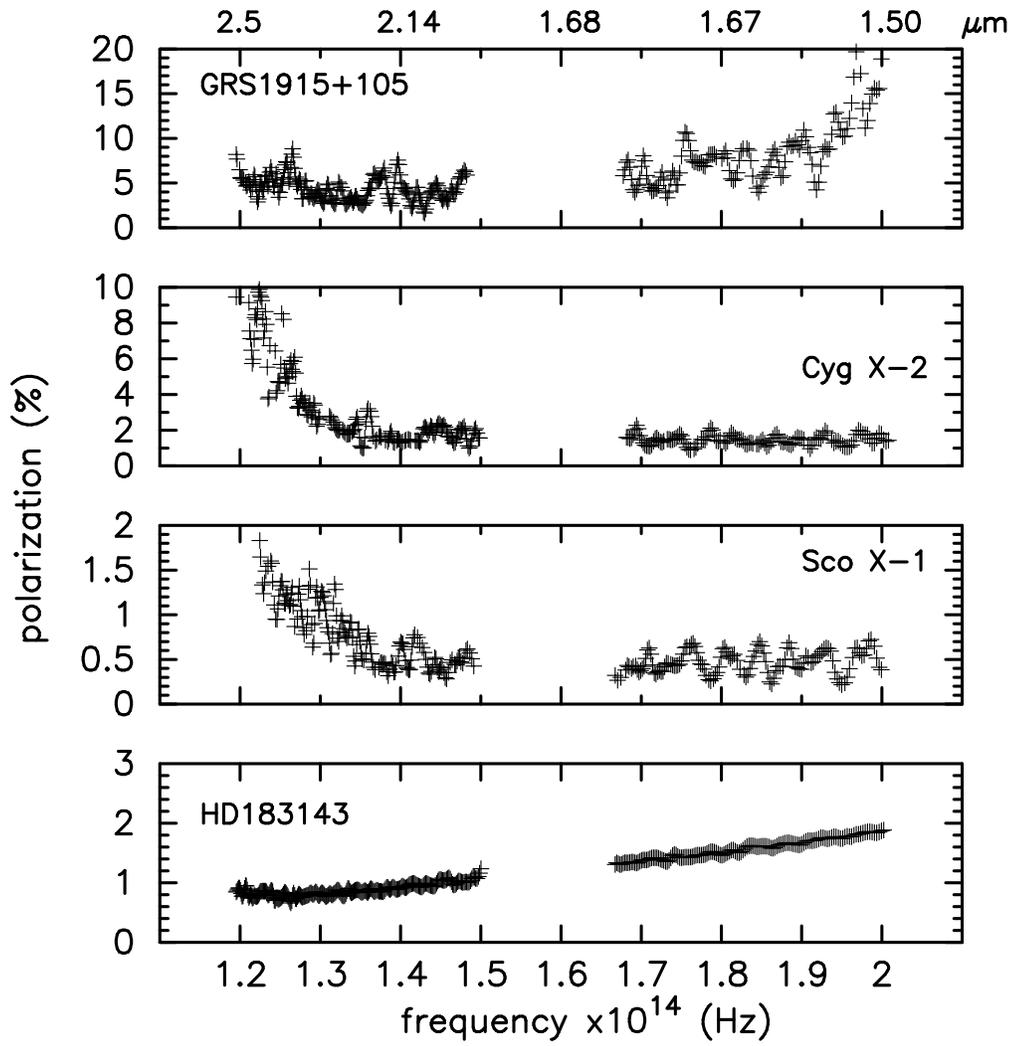}
\caption{From top to bottom: the HK linear polarization spectrum of
GRS\,1915+05,  Cyg\,X--2, Sco\,X--1 and the polarized standard star
HD184143. The gap in the spectra between 1.8 and 2.0$\mu$m is the atmospheric
absorption band.}
\label{FIG:POLSPEC}
\end{figure*}

The level and  quality of the polarization spectra were not sufficient enough 
to allow us to resolve the variations in the polarization level across the
major emission lines.  The only reliable information that could be extracted
was the continuum. In Figure\,\ref{FIG:POLSPEC} we show the wavelength binned
polarization spectrum for our targets. For Sco\,X--1 we could not determine
$\theta$ on UT 20040719 and so the polarization of the source was not detected.
Hence we only averaged the  individual Stokes $q$ and $u$ vectors on UT
20040718 and 20040723 before  calculating the polarization spectrum.   As one
can see, Sco\,X--1 and Cyg\,X--2 both show polarization spectra that increase
at longer wavelengths (lower frequencies)  i.e. the 2.4$\mu$m polarization is
larger than the 1.65$\mu$m polarization level.  For GRS\,1915+105 the opposite
is the case. In order to demonstrate that the trends observed in our science
targets are genuine, we also show the polarization spectrum of a polarized 
standard star HD183143 determined in exactly the same way as for the science
targets.  The polarization spectrum is as expected \citep{Gehrels74}. In
Table\,\ref{TABLE:OBJ} we give the polarization and position angle values.
Although the polarization standard HD183143  can be used as a validation of the
data reduction method, the standard is bright while the targets are faint,
close to the sky background level, and so poor sky subtraction may   introduce
a correlation between target brightness and polarization level. However, it
should be noted that Sco\,X--1 and GRS\,1915+105 have similar magnitudes but
very different levels of polarization.
To calculate the mean polarization at 1.65$\mu$m and 2.4$\mu$m,  we assume that
the Stokes vector do not change over 0.10$\mu$m ($N$=90 pixels).  We can thus
calculate the mean Stokes $q$ and $u$ vectors and determine the mean
polarization. The uncertainty on these values is then the standard  error,
given that we have $N$ individual  measurments of $q$ and $u$ respectively. The
uncertainty in the mean polarization  and position angle then follows by
propagating the errors on $q$ and $u$.
Given that our final mean polarization values
are  significant (note that we average over 90 pixels) a bias
correction  \citep{Simmons85} is not necessary, 
as pointed out by \citet{Jensen04}.

\subsection{Sco\,X--1}
\label{sec:scox1}

\begin{figure*}
\includegraphics[angle=-90,scale=.80]{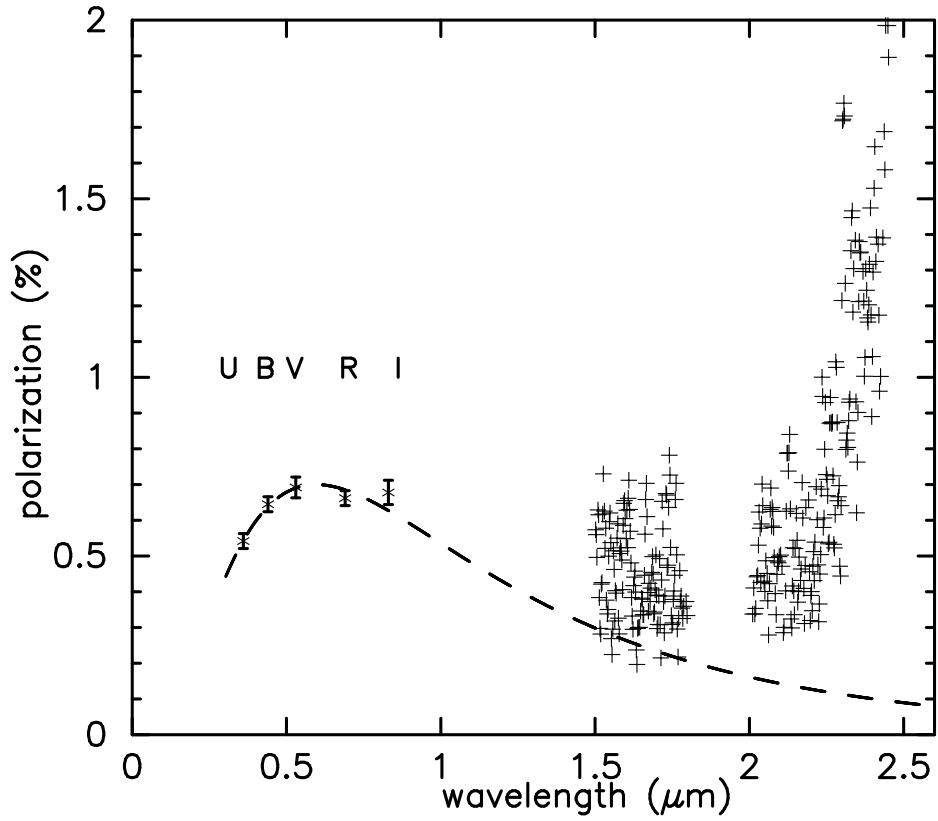}
\caption{The optical and near-infrared polarization values for Sco\,X--1. The
optical points (stars) are taken  from \citet{Schultz04} and our HK
polarization spectrum  (crosses) is shown.  The dashed line is the interstellar
polarization model fit to the optical. }
\label{FIG:ISPOL}
\end{figure*}

Sco\,X--1 has a colour excess of $E(B-V)$=0.3 so some interstellar 
polarization is expected. Schultz, Hakala, \& Huovelin (2004)  fit the UBVRI
optical linear polarization values with the empirical formula of Serkowski
($P(\lambda) = P_{\rm max} \exp^{ -K'\ln^2 (\lambda_{\rm max}/\lambda) }$; 
\citealt{Serkowski75}) and find $\lambda_{\rm max}$=596\,nm and $P_{\rm
max}$=0.70\% ($K'$=0.01+$\lambda_{\rm max}$/602). These values predict an
interstellar polarization of 0.25\% and 0.10\% at 1.65 and 2.4$\mu$m
respectively.  Figure\,\ref{FIG:ISPOL} shows the optical and near-infrared linear
polarization values and the expected interstellar polarization.  The mean
linear polarization is 0.47$\pm0.05$\% and 1.3$\pm0.10$\% at 1.65$\mu$m and
2.4$\mu$m respectively.  
Although the optical polarization may be described as interstellar, the
$H$ and $K$-band  polarization clearly cannot.  The overall optical and
near-infrared polarization spectrum can be described by two components; an
interstellar polarization spectrum in the optical, and another component which
dominates the  $H$ and $K$-band polarization  spectrum. 
It should be noted that recent broad-band JHK 
measurements of polarization of Sco\,X-1 agree with the values we obtain here
\citep{Russell07}.
Given
that Sco\,X--1 is known to be a powerful and variable jet source
\citep{Fomalont01}  the component in the IR is most likely due to 
optically thin synchrotron emission from the jet (see section\,\ref{SEC:JETS}).
We can compare the near-infrared position angles with the position angle in the 
radio images of Sco\,X--1.  The mean position angle of the radio jet on the sky is 
54 degrees \citep{Fomalont01}. This is approximately perpendicular to the near-infrared 
electric vector position angle which implies that the  magnetic field is aligned with the jet 
(assuming optically thin synchrotron emission).

\subsection{Cyg\,X--2}
\label{sec:cygx2}

Although polarization observations of Cyg\,X--2 are limited, optical linear
measurements show Cyg\,X--2 to be 0.29\% polarized, most of which may be
intrinsic to the source \citep{Miramond95}.  Our IR linear polarization
measurements of 1.7\% and 5.4\%  at 1.65$\mu$m and 2.4$\mu$m respectively show
a considerable excess in the near-infrared compared to the optical.   It is
clear that interstellar polarization cannot explain the observed optical and
near-infrared polarization values. 
The optical position angle of 36 degrees \citep{Miramond95} differs from
our near-infrared value, however, this can be explained if the jet  evolves
with time. The different position angles at  1.65$\mu$m and 2.4$\mu$m
can be explained if the different bands come from different regions. 
For a 'simple' conical jet, the jet size scales with wavelength, 
so a 2.4$\mu$m jet would be (2.4/1.65) times larger than a 1.65$\mu$m jet.
Cyg\,X--2, like Sco\,X--1, is a radio source (see \citealt{MigliariFender06})
and a member of the `Z-source' and as such is also very likely to be a jet
source (although one has never been spatially resolved).

\subsection{GRS\,1915+105}
\label{sec:grs}

GRS\,1915+105 is a powerful relativistic jet source (\citealt{Mirabel94}; 
\citealt{Fender99}; \citealt{Miller-Jones05}) and the single most convincing
example of IR synchrotron emission, with IR flare like events unambiguously
associated with similar events at millimetre and radio wavelengths (e.g.
\citealt{Fender97}; \citealt{Mirabel98}; \citealt{Eikenberry98b};
\citealt{Fender00}). However, at the time of the observations, radio monitoring
with the Ryle Telescope at 15 GHz (Guy Pooley, private communication) indicated
a low level of activity, with flux densities $\leq 10$\,mJy. 

Our near-infrared linear polarization measurements of 7.5\% and 5.0\% 
at 1.65$\mu$m
and 2.4$\mu$m may in fact be consistent with interstellar polarization, given
the very high degree of extinction  \citep{Chapuis04}. However, we can estimate
the maximum amount of interstellar polarization, given that there is a good
correlation between the interstellar extinction and  optical depth and thus
polarization at 2.2$\mu$m for field stars  \citep{Jones89}. For GRS\,1915+105
the extinction of $E(B-V)$=6.3 (or $A_v$=19.5 mags; \citealt{Chapuis04}) gives an
optical depth of $\tau_{2.2\mu m}\sim1.7$ and so a maximum interstellar
polarization of $p_{2.2\mu m}$=3.3\%. The observed polarization at 2.2$\mu$m
of  3.7$\pm1.1$\% (calculated using the mean value over 0.10$\mu$m) suggests
that the observed polarization may in fact be consistent with interstellar
polarization. 
However, if the dust induced polarization contribution is low, then
the radio jets (mean position angle of $\sim$143 degrees;  
\citealt{Miller-Jones05})  is perpendicular to the electric
near-infrared vector postion angle, implying  that the magnetic field is
aligned with the jet.

\section{Polarization from jets?}
\label{SEC:JETS}

In X-ray binaries significant linear polarization caused by electron
scattering in an isotropic medium (e.g. the environment of the
accretion flow) is expected at optical wavelengths. Indeed the linear
polarization property of the microquasar GRO\,J1655--40 in quiescence
may be explained by such a mechanism \citep{Scaltriti97}, where the
linear polarization originates from asymmetrical regions in the inner
accretion disc (electron scattering of the disc radiation by the
secondary star is insignificant, because the secondary stars in LMXBs
have low surface temperatures and therefore low free electron
densities). In the case of the persistent LMXBs where the accretion
disc dominates the optical light, the linear polarization should have
a component that is produced by electron scattering by plasma above
the accretion disc.  Although a strong interstellar component is
present, for Rayleigh and Thompson scattering processes, the level of
linear polarization decreases as a function of increasing wavelength.

An excess weak component in the optical polarization is detected in
Sco\,X--1 \citep{Schultz04}, which is most likely due to electron
scattering in the accretion disc. Similarly, for Cyg\,X--2 the optical
polarization level has been measured to be 0.28\%, which may be
intrinsic to the source. However, we find the 2.4$\mu$m linear
polarization level to be considerably larger compared to the optical
and so the 2.4$\mu$m polarization cannot be explained by interstellar
polarization or by electron scattering. In both of these sources
therefore there seems to be some intrinsic polarization of the
near-infrared which probably cannot be attributed to electron
scattering.  In the case of GRS\,1915+105 the level of polarization is
high but it does show the trend expected for interstellar scattering.

An alternative possibility is that the `excess' near-infrared polarization is due to
synchrotron emission from jets. Two polarization signatures are expected from
the jet, depending on whether or not the synchrotron emission is optically
thick or thin. Above some frequency this flat spectral component should break
to an optically thin spectrum 
($\alpha \sim -0.6$; S$_{\nu} \propto \nu^{\alpha}$) 
corresponding to the point at which the entire jet becomes transparent; i.e.
emission at the break frequency arises primarily from the `base' of the jet
(although this may be still physically separated from the central accretor).
For the optically thick part of the spectrum, which results in the flat
spectral component ($\alpha \sim 0$)   observed in the hard X-ray state 
\citep{Fender01}, no more than a few \% polarization is expected. This is
indeed observed for at least two hard state black hole sources in the radio
band (see \citealt{Fender01}) -- however such a low level of polarization would
be very hard to distinguish from the similar levels expected from scattering in
an accretion disc and/or by the interstellar medium.  

\begin{figure*}
\includegraphics[angle=-90,scale=.80]{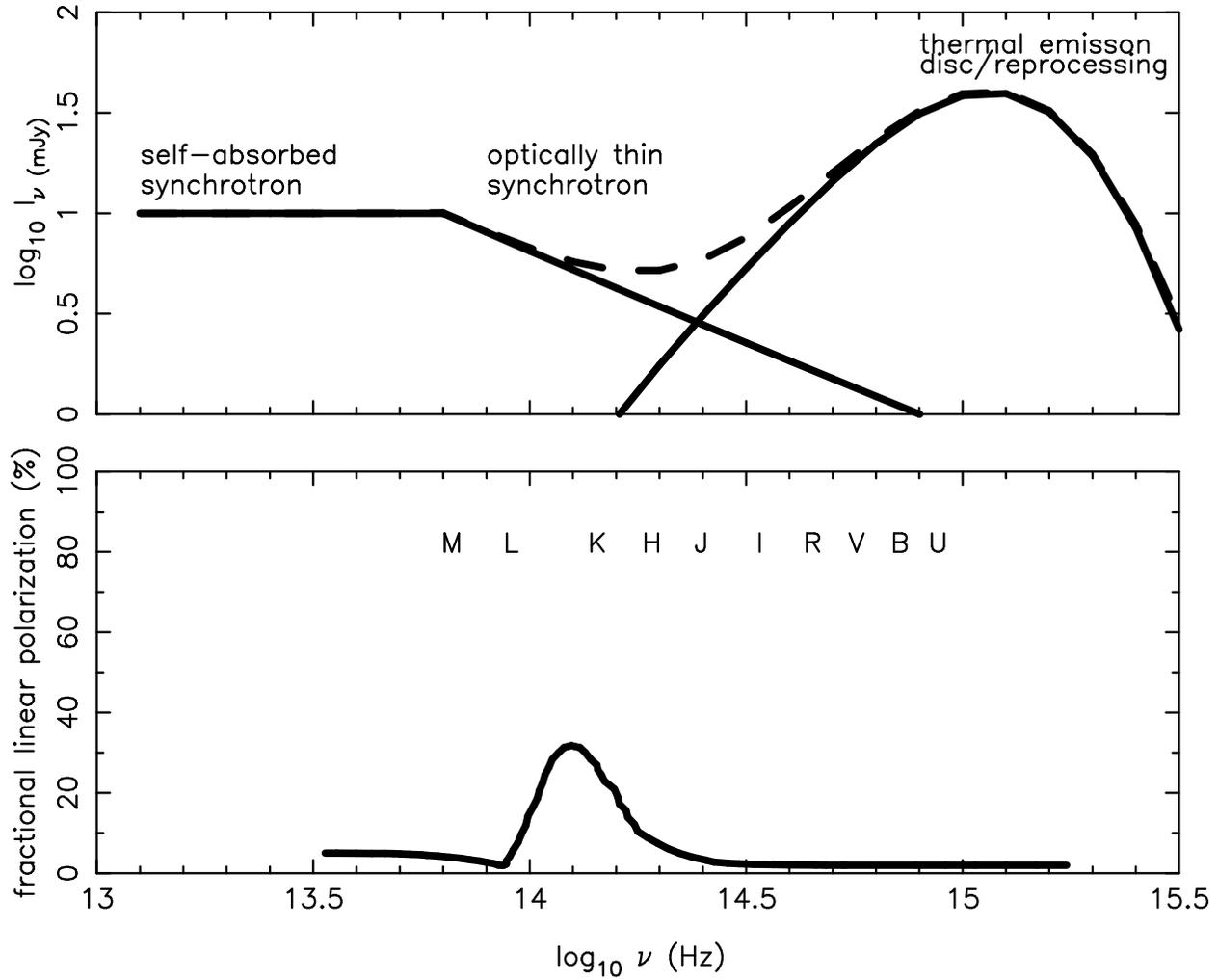}
\caption{Expectations for the linear polarization signature in the
near-infrared and optical regimes of X-ray binaries. 
The dashed line shows the observed spectral energy distribution.
The linear polarization due to optically thick of thin 
synchrotron emission is frequency dependant \citep{Pac77}.
The key spectral points are the break from optically thick (low polarization
$<$10\%)  to optically thin (high polarization $<$70\%) synchrotron emission,
and the point at which the thermal (low polarization) emission begins to
dominate over the jet. These effects should combine to produce a narrow
spectral region with relatively high linear polarization arising from
close to the base of the jet.}
\label{polfig}
\end{figure*}

There exists the possibility of a large fractional polarization level from
optically thin synchrotron emission. While it is by no means firmly
established, a small number of observational  and theoretical results suggest a
break between optically thick and thin emission should occur around the near-infrared
band (e.g. see \citealt{Corbel02}). Optically thick synchrotron emission has a
maximum linear polarization  of $\sim$10\% \citep{Blandford02}, whereas 
optically thin synchrotron emission can have a fractional linear polarization
as high as 70\% \citep{Blandford02},  where the observed level is generally
significantly less than this due to a mixing of regions with different magnetic
field orientations and/or Faraday rotations within one resolution element of
the telescope. Nevertheless, some X-ray binary jets show up to $\sim 30$\%
polarization \citep{Fender01}, indicating a highly significant ordering of the
magnetic field.  Figure\,\ref{polfig} illustrates in more detail our
expectation for the intrinsic (i.e. before interstellar scattering) linear
polarization signature in the near-infrared and optical regimes, based on spectral
energy distributions published in  \citet{Corbel02} and \citet{Homan05} and
the simple ideas outlined above about the expected polarization fraction. At
long wavelengths (maybe in the mid-infrared: the break is hard to determine
precisely) there will only be $\sim 1$\% polarization from the self-absorbed
jet; at short wavelengths a comparable level will be measured due to scattering
in the accretion flow. However, in the relatively narrow spectral region in
which optically thin synchrotron emission dominates, we may expect a strong
signature which initially rises to longer wavelengths as the relative jet:disc
fraction increases.
However, it should be noted that the break from optically thick to optically 
thin  synchrotron and the normalisation of the disc component may vary from 
source to  source, which will mean the width and peak of the high-polarization 
regime will vary. This is discussed in \citet{Nowak05} for the case of black 
hole sources and clearly illustrated by \citet{Migliari06} for the case of 
the neutron star 4U\,0614+091 where the break to optically thin synchrotron 
appears to occur  at a lower frequncy than what is measured for the black hole 
source GX\,339-4  in  \citet{Corbel02}. Therefore the sketch in 
figure\,\ref{polfig} can only be regarded as  schematic. 

Given that the 2.4$\mu$m polarization level is greater than the 1.65$\mu$m
polarization level suggests that the most likely origin of the $K$-band linear
polarization in Sco\,X--1 and Cyg\,X--2 is from a jet. Why the degree of
polarization should be different by an order of magnitude between the two
neutron star systems is unclear, but may relate to the accretion state (i.e.
branch of the `Z') the sources were on at the time (some branches seem to be
more closely associated with jets than others, e.g.  \citealt{MigliariFender06}).
These observations support the recent discovery with {\em Spitzer} of optically
thin synchrotron emission in the IR band from another neutron star X-ray
binary, 4U\,0614+091 \citep{Migliari06}.   In the case of GRS\,1915+105 the
result is ambiguous and further observations will be required to see how it
might vary in states when the jet is more powerful (as indicated by e.g.
simultaneous radio observations).

For Sco\,X--1 we find that the magnetic field is aligned with the jet 
(see section\,\ref{sec:scox1}), and the same for  GRS\,1915+105 if the dust
induced polarisation contribution is low  (see section\,\ref{sec:grs}). 
Both parallel and perpendicular magnetic fields are seen in radio jets. In the
radio  polarization maps of Cir\,X--1 one observes the polarization vector
rotate through  90 degrees between the core and the jet \citep{Fender07}. 
Parallel magnetic fields are related to the  intrinsic  jet structure, perhaps
helping to collimate the flow, whereas perpendicular  magnetic fields indicate
shocks, where the outflow is compressed along the jets axis \citep{Saikia88}.
Measurements of the linear polarization of the compact jet in the 
near-infrared  can probe closer  to the central black hole or neutron star than
the radio band, and reveal the degree or ordering of the magnetic field close
to the jets's base. Furthermore, Faraday rotation (which is proportional to
$\lambda^2$) is insignificant at near-infrared wavelengths, so that the linear
polarization electic vector position angle tells us directly about the
orientiation of the magnetic field (the two are perpendicular). The three
measurements reported here, at least two of which indicate a significant
contribution of synchrotron emission in the near-infrared, should be the
beginning of a highly useful line of enquiry in the near future.

\acknowledgments
We would like to thank Chris Davis and Brad Cavanagh for help and discussions 
with the data reduction.  We are also grateful to the referee for his/her 
careful reading of the paper. TS acknowledges support from the Spanish
Ministry  of Science  and Technology  under the programme Ram\'{o}n y Cajal.
CAW is  supported by a PPARC Postdoctoral Fellowship. The United Kingdom
Infrared Telescope  is operated by the Joint Astronomy Centre on behalf of the
U.K. Particle Physics and Astronomy Research Council. We thank the Department
of Physical Sciences, University of Hertfordshire for providing IRPOL2 for the
UKIRT.


\clearpage



\begin{table}
\begin{center}
\caption{Log of observations
\label{TABLE:LOG}}
\begin{tabular}{lllcl}
\\
\hline\hline
Date &  Object   &         UT  & Exp. time &  Comments \\
     &           &          Start &    & \\
\tableline
18 July 
&   HIP79813  & 07:48 & 8$\times$3\,s	 & telluric F0V \\	 
&   Sco\,X--1 & 08:32 & 16$\times$300\,s  &   \\
&   HD144287  & 09:22 & 8$\times$7\,s	 & non-pol stnd\\
&   HD183143  & 09:59 & 8$\times$1\,s	 & pol. stnd\\
\tableline
19 July 
&   HIP79813  & 07:48 & 8$\times$3\,s	 & telluric F0V \\	 
&   Sco\,X--1 & 08:01 & 8$\times$300\,s   &  \\
&   HD144287  & 09:18 & 8$\times$7\,s	 & non-pol. stdn\\
&   HD183143  & 09:52 & 8$\times$1\,s	 & pol. star\\
&   HIP95002      &  10:11 & 8$\times$3\,s   & telluric A2V \\
&   GRS\,1915+105 &  10:21 & 16$\times$240\,s &  \\
\tableline
22 July 
&   Cyg\,X--2 &  08:32 & 24$\times$240\,s  &   \\
&   HIP107253 &  11:01 & 8$\times$3\,s    & telluric A0V \\	  
&   HD198478  &  12:58 & 8$\times$7\,s    & non-pol. stnd\\
&   HD183143  &  12:51 & 8$\times$1\,s    & pol. star\\
\tableline
23 July 
&   HIP79813  &  05:33 & 8$\times$3\,s    &  telluric F0V \\   
&   Sco\,X--1 &  05:40 & 16$\times$300\,s  &   \\
&   HD144287  &  07:14 & 8$\times$7\,s    & non-pol. stnd\\
&   HD183143  &  07:29 & 8$\times$1\,s    & pol. stnd\\
\tableline
\end{tabular}
\end{center}
\end{table}

\begin{table}
\begin{center}
\caption{Polarimetry results. The  1.65$\mu$m and 2.4$\mu$m percentage
polarization ($p$), position angle  ($\theta$)  and dereddened flux densities
($I$) are calculated using mean values  over 0.10$\mu$m.
\label{TABLE:OBJ}}
\begin{tabular}{lllll}
\\
\hline   \hline
Object       & UT date	& 1.65$\mu$m $p$(\%) & $\theta$($^{\circ}$)   & $I$(mJy)  \\
             &          & 2.40$\mu$m $p$(\%) & $\theta$($^{\circ}$)   & $I$(mJy)  \\
\tableline
Sco\,X--1    & 20040718 & 0.38$\pm$0.04 &  136$\pm$2  & 21$\pm$2  \\
             &          & 0.93$\pm$0.05 &  147$\pm$3  & 15$\pm$1  \\
             & 20040723 & 0.67$\pm$0.04 &  116$\pm$2  & 32$\pm$2  \\
             &          & 1.14$\pm$0.06 &  129$\pm$3  & 22$\pm$2  \\
	     
Cyg\,X--2    & 20040722 &  1.7$\pm$0.2  &   96$\pm$2  & 6.4$\pm$0.4 \\
             &          &  5.4$\pm$0.7  &   84$\pm$3  & 4.3$\pm$0.2  \\
GRS\,1915+105& 20040719 &  7.9$\pm$1.0  &   49$\pm$2  & 29$\pm$3  \\ 
             &          &  5.0$\pm$1.2  &   50$\pm$3  & 15$\pm$2  \\
\tableline
\end{tabular}
\end{center}
\end{table}

\clearpage

\end{document}